\documentstyle[prl,aps,epsf,rotate]{revtex}

\newcommand{\ben}{\begin{enumerate}}
\newcommand{\een}{\end{enumerate}}
\newcommand{\be}{\begin{equation}}
\newcommand{\ee}{\end{equation}}
\newcommand{\bse}{\begin{subequation}}
\newcommand{\ese}{\end{subequation}}
\newcommand{\bea}{\begin{eqnarray}}
\newcommand{\eea}{\end{eqnarray}}
\newcommand{\bc}{\begin{center}}
\newcommand{\ec}{\end{center}}

\title{Measurements in quantum physics: towards a physical picture of relevant processes}
\author{W. A. Hofer}
\address{Dept. of Physics and Astronomy, University College London,
               Gower Street, London WC1E 6BT, UK}

\begin{document}


\maketitle

\begin{abstract}
We propose a new structure of ensembles in quantum theory, based
on the recently introduced intrinsic properties of electrons and
photons. On this statistical basis the spreading of a wave-packet,
collapse of the wave function, the quantum eraser, and
interaction-free measurements are re-analyzed and the usual
conceptual problems removed.
\end{abstract}


\section{Measurements and ensembles}
It took Albert Einstein, in 1935 \cite{einstein35}, to formulate
the basic conclusion to Heisenberg's uncertainty relations
\cite{heisenberg27}: the systems, we are dealing with in quantum
theory (QT), are not completely defined by our mathematical
representations (In case this is new to you, please read Sheldon
Goldstein's recent articles in {\em Physics Today}
\cite{goldstein98}). However, given the proof by John von Neumann
\cite{neumann32}, that the linearity of the main equations in QT
entails that no conventional ensemble can lie underneath the
statistical description of measurements, we are left with a
somewhat puzzling consequence: although there {\em must} be an
ensemble, we do not know its structure. Most lucidly, this dilemma
has been expressed by David Bohm \cite{bohm66a}:

{\em Yet it is not immediately clear, how the ensembles, to which
... probabilities refer, are formed and what their individual
elements are. For the very terminology of quantum mechanics
contains an unusual and significant feature, in that what is
called the physical state of a quantum mechanical system is
assumed to manifest itself only in an ensemble of systems.}

By now the way Bohm proposed to amend this deficiency in the
logical structure of QT is well researched: if the wavefunction at
a given location and moment is a complex number defined by

\be
\psi = R \cdot e^{iS} \ee

\noindent then the Schr\"odinger equation gives rise to a
''quantum potential''

\be
Q = \frac{\nabla^2 \, R}{R} \ee

\noindent which, in the same way a physical potential $V$
determines the solutions of Schr\"odinger's equation, determines
the mathematical form of the wavefunction $\psi$. There is an
important difference, though: since $Q$ is proportional to the
second derivative of $R$, the value of $\psi$ at one point of the
system influences $Q$ throughout the whole system and vice versa:
the very ansatz of Bohm thus gives rise to strongly non-local
effects \cite{holland93}.

This non-locality of Bohm's theory was, initially, taken as an
argument against its theoretical soundness (see Goldstein's
article), until John Bell and Alain Aspect proved that the same
applies to QT itself \cite{bell64,aspect82}. Since then the
community is divided: while the more orthodox faction believes in
{\em physical} non-locality regardless of the contradiction with
Special Relativity (see e.g. the recent article in {\em Nature} on
''quantum teleportation'' \cite{bouwmeester97}), the more cautious
view is that we are dealing with a {\em conceptual} non-locality
due to our representation of micro physical systems. It seems that
only a clear perception of the ensemble structure in QT will allow
a decision in this issue. As previous work revealed, the features
of the ensembles in quantum theory could arise from intrinsic
fields due to particle propagation \cite{hofer98a}. This approach
goes beyond the proposed modifications of Everett
\cite{everett57}, Ghirardi \cite{ghirardi86}, Omnes
\cite{omnes92}, Hartle \cite{hartle91}, Griffiths
\cite{griffiths84}, Zurek \cite{zurek82} and others, who, although
pioneering new ways to solve the measurement problem in QT,
confined themselves more or less to the question of wavefunction
collapse.

The strategy employed in developing this new framework was roughly
the following: since the mathematical description in QT yields
correct results, it should be feasible to complete these -
mathematical - expressions by a physical basis. It is evident that
the conventional framework of QT allows for no such basis, since
the wavefunction has no physical meaning. In the new framework, on
the contrary, the wavefunction gains a double meaning: a physical
as well as a statistical one, corresponding to the description of
single particles (single elements in Bohm's words) and ensembles
of particles. The argument developed in the present paper will be
that this double meaning is responsible for some of the most
puzzling results in quantum theory. To see, where this double
meaning comes from, let me briefly sketch the development of the
theoretical framework, for a more thorough view please refer to
the original paper \cite{hofer98a}.

\section{Theoretical foundations}\label{meas_theory}

That the energy of a particle depends on its {\em intrinsic}
frequency $\omega$ is the basis of Planck's quantum hypothesis,
thus:

\begin{equation}
E = \hbar \cdot \omega
\end{equation}

For photons, the frequency is equal to the frequency of its
electromagnetic $\vec{E}$ and $\vec{B}$ fields. For massive
particles like electrons there is no such connection, one either
has to refer to de Broglie's {\it harmony of phases}
\cite{debroglie25}, or seek a solution in terms of intrinsic and
rotational components of motion (e.g. in \cite{recami97}).
However, if the wave equation describes the intrinsic density of
mass distribution $\rho = \rho(\vec{r},t)$, oscillating with the
characteristic frequency $\omega$, then the total energy of the
electron is split into two distinct components \cite{hofer98a}:

\be
E  =  \hbar \cdot \omega =: m_{e} u^2 \qquad E_{QT}  = E_{K} =
\frac{m_{e}}{2} u^2 \qquad E_{F} = \frac{m_{e}}{2} u^2 \ee

where $u$ is the velocity, $m_{e}$ the mass, and $E_{QT} = E_{K}$
the {\em kinetic} energy of the electron, equaling the expression
in quantum theory. $E_{F}$, the {\em field} energy results from
the intrinsic wave features of electron motion. While the latter
is, initially, purely hypothetical, it can be justified, a
posteriori, by calculating the solutions to Schr\"odinger's
equation under the condition that $E_{F}$ in QT is generally
undefined. From the time-independent equation:

\begin{equation}
\left( - \frac{\hbar^2}{2 m_{e}} \nabla^2 + V \right) \psi = E
\cdot \psi
\end{equation}

and in a plane wave basis set the undefined field component of
electron energy leads to an uncertainty in the range of the
wavevector $\vec{k}$, and the consequence of this uncertainty can
be expressed in Heisenberg's uncertainty relations
\cite{hofer98a,heisenberg27}:

\be
\Delta P \, \Delta X \ge \frac{\hbar}{2} \ee

If the field energy $E_{F}$ is described by an intrinsic scalar
field $\phi_{F}$ and vector fields $\vec{E}$, $\vec{B}$ so that:

\be
\phi_{F}  =  \frac{1}{2} \left( \frac{1}{u^2} \, \vec{E}^2 +
\vec{B}^2 \right) \qquad E_{F} = \int_{V} dV \phi_{F}  \qquad
\vec{E} \cdot \vec{u}  =  \vec{B} \cdot \vec{u} = \vec{E} \cdot
\vec{B} = 0 \ee

then it can be established that these $\vec{E}$ and $\vec{B}$
fields comply with the Maxwell equations
\cite{hofer98a,jackson84}. Please note that the shape of the
particles or their volume do not enter the picture, they remain
completely undefined. In addition, all the conventional Planck and
de Broglie relations are valid, although the latter refer to phase
velocity rather than group velocity. Based on these intrinsic
features the {\em physical} meaning of the wavefunction is that
the square of its real part $\Re^2 (\psi)$, for a single particle,
is proportional to its density of mass $\rho$. The statistical
meaning is different for electrons and photons, since only the
electron has an undefined intrinsic energy component.

In the following we shall simplify the model by assuming that the
density of mass of any single particle is constant for a specific
type (electron, photon), and described by $\rho = |\psi|^2$. The
undefined field component of energy $E_{F}$ shall be accounted for
by a range of particle energies rather than its development with
time. The model treated in the next sections is therefore a static
account of an essentially dynamical problem. However, as will be
seen presently, this simplified account is sufficient to shed new
light on some of the most interesting problems in measurement
theory. The statistics in measurement processes result from the
following unknown variables:

\begin{itemize}
\item
{\bf Photons:} unknown phase, unknown amplitude of wavefunction,
unknown number of particles.

\item
{\bf Electrons:} unknown phase, unknown amplitude of wavefunction,
unknown intrinsic energy, unknown number of particles.

\end{itemize}

\section{Quantum ensemble of free electrons}

Due to periodic wave functions, the intrinsic potential at a
moment $t$ can take any value, and the wave vector of the problem
therefore is not exactly determined, but covers the whole range
from $k^2 = 0$ to $k^2 = \frac{m}{\hbar^2} E_{T}$, where $E_{T} =
\hbar \omega$. Consider a point $\vec r$, where the external
potential vanishes $V(\vec r) = 0$. Due to the disregard for
intrinsic potentials the Schr\"odinger equation at this location
applies for all wavelets described by:

\be\label{ar001} \vec k^{2}(\vec r) + \vec k_{i}^2(t) =
\frac{m}{\hbar^2} E_{T} \qquad 0 \leq \vec k_{i}^2(t) \leq
\frac{m}{\hbar^2} E_{T} \ee

Here $ \vec k_{i}^2(t) $ shall denote the intrinsic potential, not
accounted for in QM. $E_{T}$ is the total energy of the particle.
The two variables are given by:

\be\label{ar002} E_{T} = m u^2 \qquad \frac{\hbar^2}{m}\vec
k_{i}^2(t) = \phi_{i}(t) V_{p} \ee

where $u$ again is the velocity and $V_{P}$ the volume of a
particle. The {\em quantum ensemble} shall be the integral over
allowed wave states. Then the wave function $\psi(\vec r)$ can be
written as:

\be\label{ar003} \psi(\vec r) =  \frac{1}{(2 \pi)^{3/2}}
\int_{0}^{k_{0}} d^3k \, \chi_{0}(\vec k) e^{i \vec k \vec r}
\qquad k_{0} = \sqrt{\frac{m}{\hbar^2}E_{T}} \ee

\noindent Using a Fourier transformation the amplitudes of the
ensemble are:

\be\label{ar004} \chi_{0}(\vec k) = \frac{1}{(2 \pi)^{3/2}} \,
\int_{-\infty}^{+\infty} d^3 r\, \psi(\vec r) e^{- i \vec k \vec
r} \ee

In this case an ensemble of electrons is described by identical
solutions of the Schr\"odinger equation, although in QT the
mathematical representation seems to refer to one particle and one
particle only. Bohm's puzzlement with the {\em individual system},
which is in fact an {\em ensemble of systems} then is quite
justified.

\section{Quantum ensemble in external potentials}

An even more interesting consequence of the same feature is
observed in external potentials. The external potential at a point
$\vec r$ has two effects: the range of allowed wavelets and
therefore the quantum ensemble will be changed, and the internal
properties of single wavelets will be altered. If the potential at
$\vec r$ equals $V(\vec r)$, the allowed $k$--values will comply
with:

\be\label{ar007} k^2(\vec r) + k_{i}^2(t) = \frac{m}{\hbar^2}
\left(E_{T} - V(\vec r) \right) \qquad 0  \leq  \vec k_{i}^2(t)
\leq \frac{m}{\hbar^2} \ee
\be
\left(E_{T} - V(\vec r) \right) k_{1}^2 = \frac{m}{\hbar^2}(E_{T}
- V(\vec r)) \quad E_{T} = m u^2 \nonumber \ee

For reasons of consistency the potential $V(\vec r)$ is double the
potential if only kinetic properties are considered. The range of
allowed $k$--values in this case depends on the energy $E_{T}$ of
a single particle as well as the potential applied. There are two
distinct cases: $E_{T} - V(\vec r)$ is either a positive or a
negative value, corresponding to wavelets in a potential or to
exponential decay of single waves.

For $E_{T} - V(\vec r) > 0$ the potential can either be a positive
or a negative value, leading to an enhancement or a reduction of
the quantum ensemble of valid solutions. The general solution for
both cases is then:

\be\label{ar008} \psi(\vec r) =  \frac{1}{(2 \pi)^{3/2}}
\,\int_{0}^{k_{1}} d^3k \, \chi_{0}(\vec k) e^{i \vec k \vec r}
\nonumber \\ k_{1} = \sqrt{\frac{m}{\hbar^2}(E_{T} \pm |V(\vec
r)|)} \quad E_{T} = m u^2 \ee

\begin{figure}
\epsfxsize=1.0\hsize \epsfbox{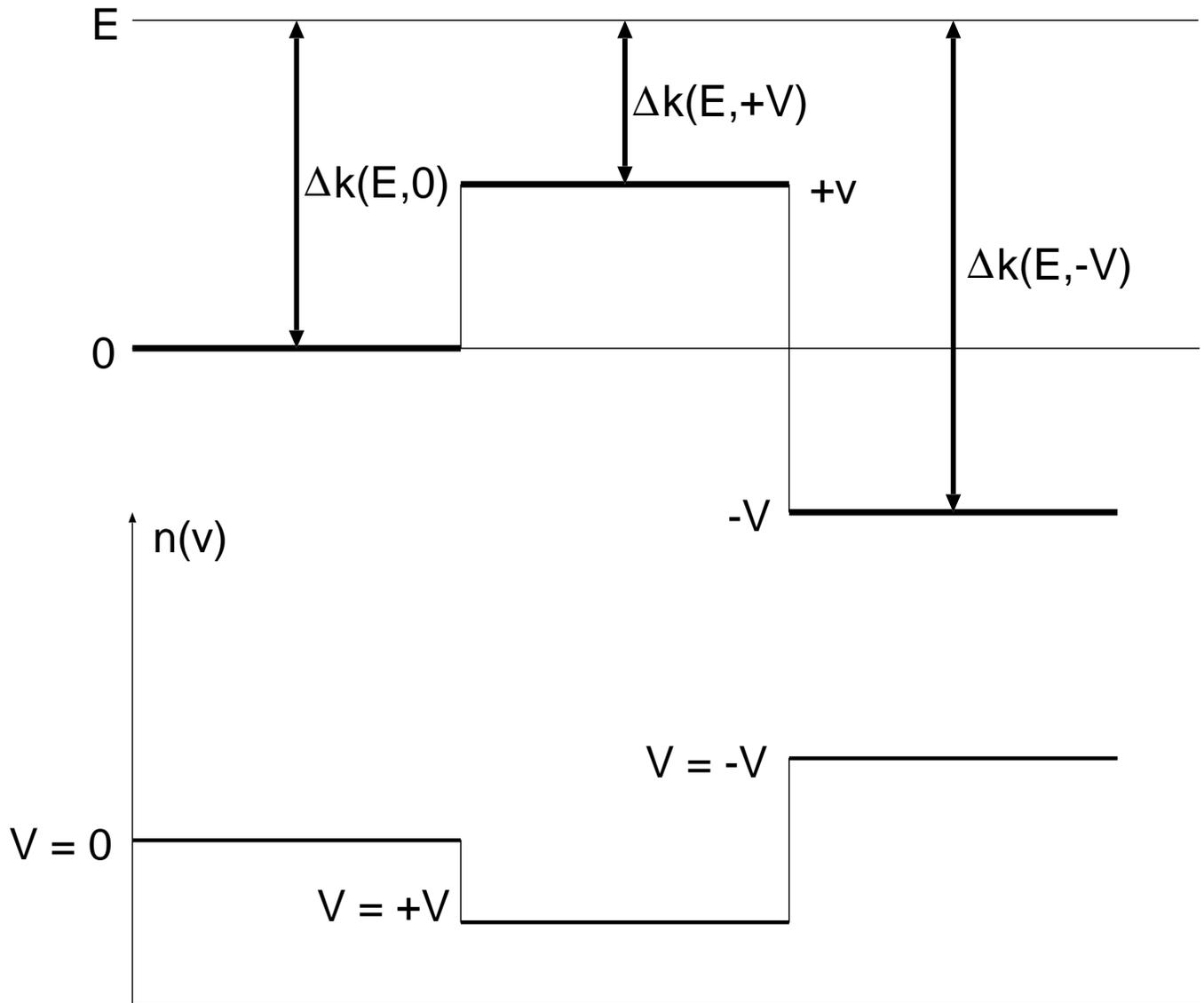}
\caption{Quantum
ensemble of wavelets for three different potentials. A negative
potential increases the number of allowed $k$--values, a positive
potential has the opposite effect}\label{fig001}
\end{figure}

The range of individual wavelets defines, as in the case of a
vanishing external potential, an ensemble of particles, which
comply with the differential Schr\"odinger equation. Any
integration of the equation therefore also contains a -- hidden --
manifold of individual wavelets. A positive potential essentially
limits the number of individual waves contained in the ensemble,
because it diminishes the range of $k$. A negative potential has
the opposite effect: the number of waves in the ensemble is
increased, and their statistical weight in the whole system is
equally higher. \mbox{Fig. \ref{fig001}} displays the quantum
ensembles for different external potentials.

For energies $E_{T} < V(\vec r)$ the mathematical formalism of
Schr\"odinger's equation allows for solutions with a negative
square of $\vec k$, equivalent to an exponential decay of single
wavelets:

\be\label{ar009} k^2(\vec r) + k_{i}^2(t) = \frac{m}{\hbar^2}
\left(E_{T} - V(\vec r) \right) \le 0 \qquad 0  \geq \vec
k_{i}^2(t) \geq \frac{m}{\hbar^2} \left(E_{T} - V(\vec r) \right)
\ee
\be
\psi(\vec r) = \frac{1}{(2 \pi)^{3/2}} \,\int_{0}^{k_{1}} d^3k \,
\chi_{0}(\vec k) e^{- \vec k \vec r} \qquad  k_{1} =
\sqrt{\frac{m}{\hbar^2}(V(\vec r) - E_{T})} \quad E_{T} = m u^2
\ee

The question in this case concerns not so much the mathematical
formalism but the physical validity. Considering, that
electrodynamics is a description of intrinsic properties of single
particles, applicable to photons as well as electrons, the results
of electrodynamics in different media should also have relevance
for the wave properties of single particles. And considering,
furthermore, that an exponential decay into a medium at a boundary
is one type of solution, the same must generally hold for single
wavelets. It is a basically {\em classical} solution to boundary
value problems and, so far, no specific feature of a quantum
system.

\section{Wave function normalization}

We have not yet defined the amplitude $\chi_{0}(\vec k)$ of single
components in the quantum ensemble. This can be done by requiring
single wavelets to comply with the mass relations of particles.
Since:

\be\label{ar010} \chi^{*}(\vec k) \chi(\vec k') = \frac{1}{(2
\pi)^3} \chi_{0}(\vec k) \chi_{0}(\vec k') e^{i \vec r (\vec k' -
\vec k)} \ee

\noindent an integration over infinite space yields the result:

\be\label{ar011} \int_{-\infty}^{+\infty}d^3 r\,\chi^{*}(\vec k)
\chi(\vec k') = \delta^3(\vec k - \vec k') \chi_{0}(\vec k)
\chi_{0}(\vec k') \qquad m = \int_{-\infty}^{+\infty}d^3
r\,|\chi(\vec k)|^2 = \chi_{0}^2(\vec k) \ee

Using this amplitude the square of the wave function $\psi(\vec
r)$ in different external potentials can be calculated.

\bea\label{ar012} \int_{-\infty}^{+\infty}d^3r\,|\psi(\vec r)|^2
&=& \frac{m}{(2 \pi)^3} \int d^3r \int_{0}^{k}d^3k d^3k'e^{i \vec
r (\vec k' - \vec k)} \nonumber \\
 &=&  m \int_{0}^{k}d^3k\,d^3k'
\delta^3(\vec k - \vec k') = \frac{4 \pi m}{3}\,k^3 \eea

If we consider a system, where the potential $V = V(\vec{r})$, the
wavefunction will depend, via the normalization condition, on the
potentials in all parts of this system. And if we formalize two
space-like separated events within a single system in QT (there is
no way to avoid this in any experiment aimed at proving or
refuting non-locality), these events are no longer logically (or
physically) separate: non-locality is therefore an inherent
conceptual feature of quantum theory.

\section{Boundary conditions}

One of the easiest examples in quantum theory, which suffices to
demonstrate the effect of boundary conditions, is the square
potential well. We take a one dimensional potential well, the
external potentials described by:

\be \label{ar102} V = 0  \quad \forall \,|x| \le x_{0} \qquad V =
V_{0}  \quad \forall \,|x| \ge x_{0} \ee

To solve the problem we have two consider the behavior of single
members of the quantum ensemble as well as the behavior of the
whole ensemble. The limiting $k$ values can again be inferred from
the solution of the one--dimensional Schr\"odinger equation, they
will be for $E_{T} < V_{0}$:

\begin{eqnarray} \label{ar103}
k_{0}^2 &\le& \frac{m}{\hbar^2} E_{T} \quad  |x| \le x_{0}
\nonumber \\ k_{0}'^2 &\le& \frac{m}{\hbar^2} (V_{0} - E_{T})
\quad  |x| \ge x_{0}
\end{eqnarray}

For incident, reflected and penetrating waves, the three
components of the wave are a wave of positive and a wave of
negative propagation in the region $ |x| < x_{0}$, and a decaying
component in the region $|x| > x_{0}$. Considering individual
members of the ensemble, the lowest $k$ value should correspond to
maximum decay in the potential, while the member with maximum
total energy should display maximum penetration. The relation
between an arbitrary wave vector $k_{1}$ and its corresponding
member $k_{2}$ must therefore be:

\be\label{ar104} k_{1}^2 + k_{2}^2 = \frac{m}{\hbar^2} V_{0} \ee

The wave functions in the three separate regions shall be
described by standard solutions. Accounting for  the boundary
conditions for steady transition of the particle wave the
coefficients can be determined and the solution for an individual
wave is therefore, equivalent to the solution in quantum theory
\cite{COH77}:

\be\label{ar105} \qquad \qquad \chi_{0} \cdot e^{k_{2} x}  \qquad
\qquad \qquad x \le - x_{0} \ee
\be
\chi(x) = \chi_{0} \cdot e^{- k_{2} x_{0}} \frac{\cos k_{1}
x}{\cos k_{1} x_{0}} \qquad |x| \le x_{0} \ee
\be
\qquad \qquad \chi_{0} \cdot e^{- k_{2} x} \qquad \qquad \qquad x
\ge x_{0} \ee

\noindent The normalization condition an ensemble member yields
the amplitude of the particle wave:

\begin{eqnarray}\label{ar106}
\chi_{0}(k_{1},k_{2}) = \sqrt{\frac{m k_{2}}{1 + k_{2}
x_{0}}}e^{k_{2}x_{0}} \cos k_{1} x_{0}
\end{eqnarray}

\noindent And the total ensemble can equally be calculated by
integrating over the full range of allowed $k$ values.

\bea\label{ar107} \int_{\infty}^{\infty} dx |\psi(x)|^2 &=& 2
\int_{0}^{\infty} dx \left[ \theta(x_{0}-x) \int_{0}^{k_{0}}
dk_{1}\, \chi_{0}^2(k_{1})  e^{- 2 k_{1} x_{0}} \frac{\cos^2(k_{1}
x)} {\cos^2(k_{1} x_{0})} \quad + \right. \\ &+& \left.
\theta(x-x_{0}) \int_{0}^{k_{0}'} dk_{2}\, \chi_{0}^2(k_{2}) e^{-
2 k_{2} x} \right] \nonumber \eea

The amplitude $\chi_{0}(k)$ must finally be renormalized, and the
square of the wave function then describes the probability
distribution of the whole ensemble. The procedure described is
similar to the standard procedure in quantum theory, although in
this model $k$ space is not unlimited, the cutoff is determined by
the energy $E_{T}$. The structure of the ensembles in different
environments provides a means to analyze the interplay between
physical aspects (the intrinsic properties of single electrons)
and statistical ones (the way quantum theory contains a hidden
ensemble of single electrons).

\section{Spreading of a wave--packet}

The effect is a common source of irritation, and concepts have
been put forth to eliminate it in a modified version of quantum
theory (see, for example, Mackinnon \cite{MAK78,DAT85}, or the GRW
model \cite{ghirardi86}). In a one dimensional model the
development of an amplitude $\hat\psi(k)$ is described by the
integral:

\be\label{ar201} \psi(x, t) = \int dk \,e^{i(kx - \omega t)}
\hat\psi(k) \ee

Two initial distributions are calculated: a Gaussian distribution,
with a wave function centered around a value $x_{0} = 0$ at $t =
0$, and an ensemble with exactly defined energies (such an
ensemble can be obtained by energy measurements, as shown further
down):

\be\label{ar202} \psi_{1}(x, t = 0) = e^{- \frac{x^2}{b^2} + i
k_{0}x} \qquad \psi_{2}(x, t = 0) = e^{i k_{0} x} \ee

\noindent The evaluation of the integral yields:

\bea\label{ar203} \hat\psi_{1}(k) = \int dx \,\psi_{1}(x,0)\,e^{-
i k x} = e^{-(k - k_{0})^2\,b^2/2} \nonumber \\ \hat\psi_{2}(k) =
\int dx \,\psi_{2}(x,0)\,e^{- i k x} = \delta (k - k_{0}) \eea

\noindent The square of the two wave functions at $t > 0$ is
consequently:

\be\label{ar205} \left| \psi_{1}(x, t > 0) \right|^2 = \left( 1 +
\frac{\hbar^2t^2}{m^2 b^4}\right)^{-1}
 \exp \left[ - b^{-2}
\left( 1 + \frac{\hbar^2t^2}{m^2 b^4}\right)^{-2} \cdot \left( x -
\frac{\hbar k_{0}}{m} t \right)^2 \right] \ee
\be
\left| \psi_{2}(x, t > 0) \right|^2 = 1 \ee

\begin{figure}
\epsfxsize=1.0\hsize \epsfbox{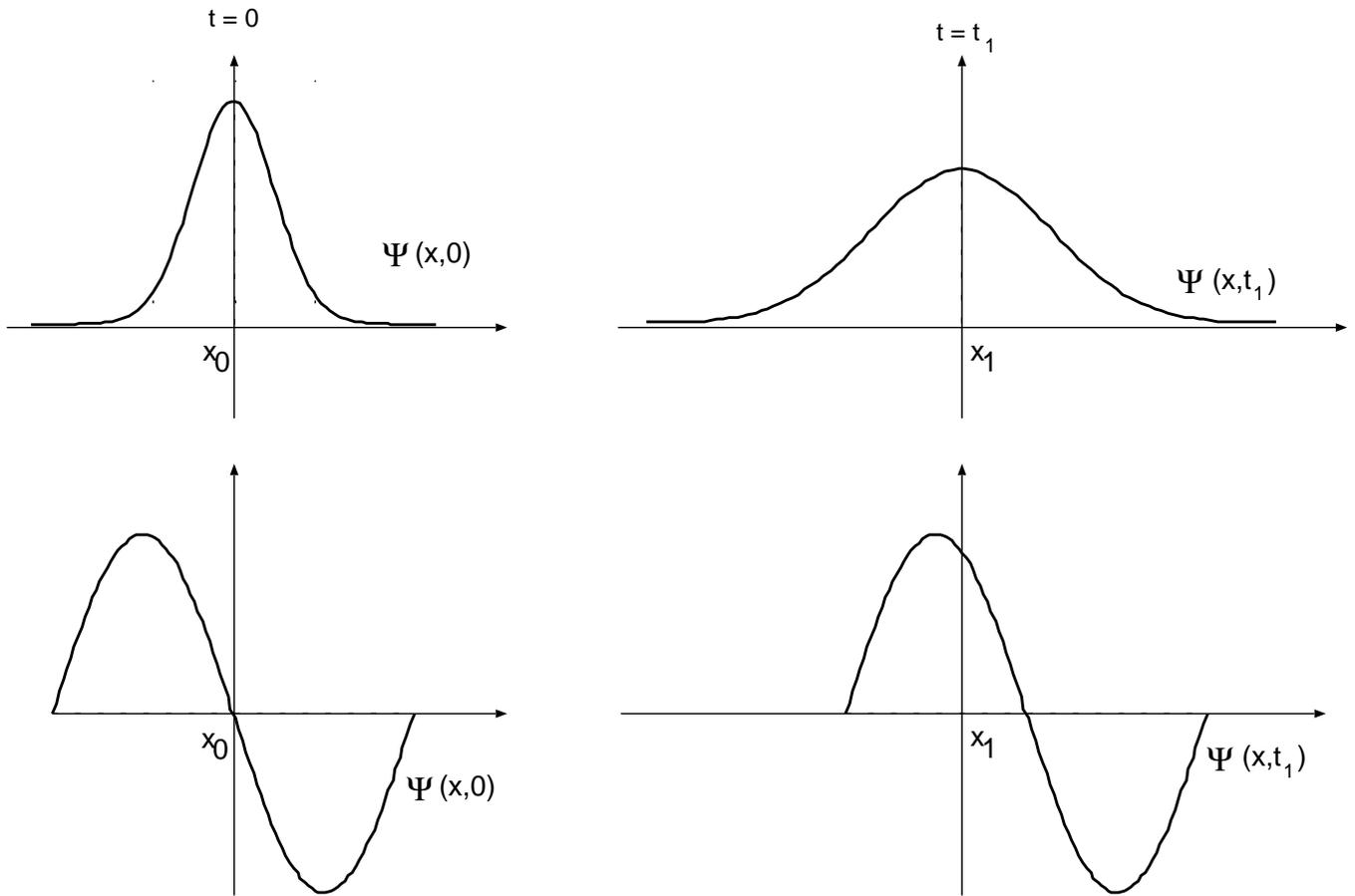}
\caption{Development of
ensembles. An ensemble of arbitrary energy and Gaussian
distribution will develop into a quantum ensemble of equal
probability for all energy values (top). The development of a
particle with exactly defined energy due to the Schr\"odinger
equation or the wave equation leads to a local ensemble over the
whole local range of the system (bottom)}\label{fig002}
\end{figure}

The development of the two ensembles is shown in Fig.
\ref{fig002}. As it is possible to decompose an arbitrary
distribution of initial $k$ values in Gaussian distributions, the
result holds quite generally. There are two possibilities to
account for this feature: (i) Either the restructuring is referred
to some potential not covered by field theories, in this case we
would have to recur to Bohm's quantum potential \cite{BOH52}. (ii)
The initial conditions contain an assumption which affects the
physical properties of particles. In the single particle case (see
section \ref{meas_theory}), where $[\Re \, \psi(x)]^2 \propto
\rho(x) \propto \phi(x)$, an inhomogeneous distribution like $
\psi_{1}(x, 0)$ gives rise to an intrinsic potential $\phi(x)$,
described by:

\be\label{ar206} \psi_{1}(x,0) = \psi_{0} e^{i k_{0} x} \qquad
\psi_{0} = e^{- x^2/2b^2} \nonumber \\ \phi(x, 0) = \psi_{0}^2 =
e^{- x^2/b^2} \ee

And the system is therefore not, as implied by the mathematical
formulations, free of forces, but will experience a force along
the direction $x$:

\be\label{ar207} F_{x} = - \frac{\partial \phi}{\partial x} =
\frac{2 x}{b^2} e^{- x^2/b^2} \ee

Along this line of reasoning we may reconsider the question of
quantum ensembles from the viewpoint of intrinsic potentials and
forces. The most general form of a wave function is given by the
integral:

\be\label{ar208} \psi(\vec r) = \int d^3k \, \psi_{0, \vec k}(\vec
r, \vec k) e^{i \vec k \vec r} \ee

The potentials due to the qualities of the amplitude are then
responsible for intrinsic energy components in addition to the
purely periodic fields of a plane wave. They are:

\be\label{ar209} \phi_{i}(\vec r, \vec k) = u^2 \, \left|\psi_{0,
\vec k}(\vec r, \vec k)\right|^2 = \frac{\hbar^2 k^2}{m^2}
\left|\psi_{0, \vec k}(\vec r, \vec k)\right|^2 \ee

The forces within the propagating wave are either periodic -- the
total energy density of the plane wave is constant --, or they are
forces due to the properties of the amplitude. These forces will
be:

\be\label{ar210} \vec F = \frac{\hbar^2 k^2}{m^2} \left[ \psi_{0,
\vec k}^{*}\, \nabla \psi_{0, \vec k} + \psi_{0, \vec k} \, \nabla
\psi_{0, \vec k}^{*} \right] \ee

A stable state of the system can only be expected, if these forces
vanish. The equilibrium condition for a system of particles
described as plane waves is therefore:

\be
\psi_{0, \vec k}^{*} \, \nabla \psi_{0, \vec k} + \psi_{0, \vec k}
\, \nabla \psi_{0, \vec k}^{*} = 0 \ee

The two ensembles defined, the ensemble with exact energies as
well as the quantum ensemble (equal amplitude of all partial
waves), comply with this condition since in both cases the
amplitudes $ \psi_{0, \vec k}$ do not depend on $\vec r$. But the
distribution used for the calculation of the spreading wave packet
is incompatible with this condition.

\section{Collapse of the wave function}\label{meas_co}

In his rather fundamental and comprehensive analysis of
measurement processes in quantum theory Ballentine
\cite{BAL70,BAL84} proceeded from two mutually exclusive
statements on the quality of the state concept, i.e. that (i) {\em
a pure state provides a complete and exhaustive description of an
individual system}, and (ii) {\em a pure (or mixed) state
describes the statistical properties of an ensemble of similarly
prepared systems}. The subsequent analysis of measurement
processes proved that \cite{BAL84} {\em any interpretation of the
type (i) \ldots is untenable}.

\begin{figure}
\epsfxsize=1.0\hsize \epsfbox{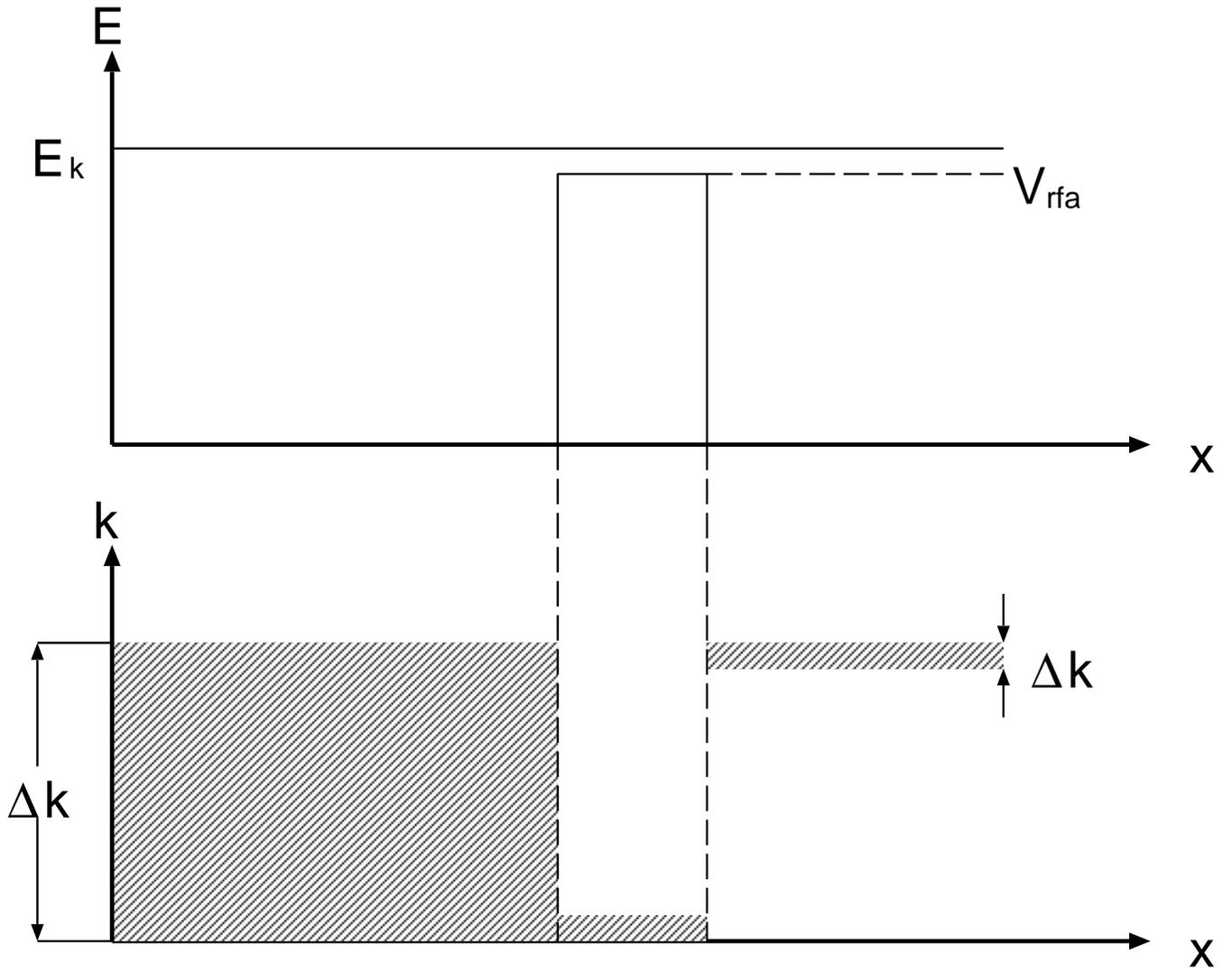} \vspace{0.5cm}
\caption{Reduction of the wave function due to retarding field
analyzer. The positive potential $V_{rfa}$ is lower than the total
energy of the ensemble limit $E_{k}$ (top), the selection of
members of the ensemble leads to a reduction of the statistical
ensemble after the measurement (bottom) }\label{fig003}
\end{figure}

If the state vector of a system, or the wave function in QT were
an exhaustive information about the system, the logical problems
seem indeed severe if not unsurmountable. The situation changes,
though, if one concedes that the quantum mechanical description
does not provide a full account of physical variables. To a
greater or lesser extent all the proposed modifications of QT to
account for the measurement problems, cited in the introduction,
use this feature. As a simple example of the reduction of the wave
function in a measurement process we consider a retarding field
analyzer frequently employed in LEED (low energy electron
diffraction) measurements. A retarding field analyzer is a
positive potential, assumed rectangular for simplicity, which
selects only electrons above a certain energy threshold. We
equally assume, that the electrons initially are free, their
energy shall be given by a value $E_{k}$ (see Fig. \ref{fig003}).
From a strictly causal point of view, the electrons below an
exactly defined threshold value $E_{rfa} < E_{k}$ cannot pass the
filter and the number of electrons after the filter is therefore
reduced to single particles with an energy above the threshold
value. As the calculation in quantum theory integrates over all
possible particle states at a given location $\vec r$, and since
the range of allowed $k$--values depends on the level of kinetic
energy, the total density $\rho(\vec r)$ at any location before
the analyzer will be:

\be\label{if011} \rho(\vec r) = \psi^{*}(\vec r) \psi(\vec r) =
\int_{0}^{k_{0}}d^3 k \chi^{*}(\vec k,\vec r) \chi(\vec k,\vec r)
\nonumber \\ k_{0}^2 = \frac{2 m}{\hbar^2}\,E_{k} \ee

while after the analyzer the wave function will be limited to
states with energy values higher than the threshold:

\be\label{if012} \rho'(\vec r) = \psi'^{*}(\vec r) \psi'(\vec r) =
\int_{k_{1}}^{k_{0}}d^3 k \chi^{*}(\vec k,\vec r) \chi(\vec k,\vec
r) \nonumber \\ k_{1}^2 = \frac{2 m}{\hbar^2}\,(E_{k} - E_{rfa})
\ee

Clearly the quantum ensemble has been reduced. The measurement of
particle energies by retarding fields therefore leads to a
reduction of the statistical ensemble, but in a causal and
deterministic manner. The ensemble wave function is reduced due to
the removal, in an equally causal and deterministic way, of
partial waves. It should be noted that the collapse of the wave
function in real space cannot, presently, be treated within the
same model.

\section{The quantum eraser}

The polarization of the intrinsic fields is decisive for
interference measurements as can be demonstrated by an analysis of
{\em quantum eraser} phenomena. In this case the which--path
information of the photon is said to preclude interference.
Conventionally, the measurement is formalized as follows
\cite{SCU82}. The amplitude of an incident photon of horizontal
polarization is split coherently in two separate beams, described
by the quantum state vector ($1$ denotes the first, $2$ the second
of the two paths)

\be
\psi_{12}^0  = \frac{1}{\sqrt{2}} \left( \psi_{1,H} + \psi_{2,H}
\right) \ee

\noindent The square of $\psi$, or the probability density in this
case contains an interference term $\psi_{1,H}^{*} \psi_{2,H}^{
}$:

\[
|\psi_{12}^0|^2 = \frac{1}{2} \left( |\psi_{1,H}|^2 +
|\psi_{2,H}|^2 + \psi_{1,H}^{*} \psi_{2,H} + \psi_{2,H}^{*}
\psi_{1,H} \right)
\]

If a polarization rotator changing the polarization of the beam to
vertical (V) orientation is placed in path 1, the interference
pattern is no longer observable, and the measurement yields random
results for the local probability density on the measurement
screen. In quantum theory the result is referred to the
orthogonality of the two states $H$, and $V$, and the state vector
of the photon described by:

\be
\psi_{12}^1 = \frac{1}{\sqrt{2}} \left( \psi_{1,V} + \psi_{2,H}
\right) \nonumber \\ |\psi_{12}^1|^2 = \frac{1}{2} \left(
|\psi_{1}|^2 + |\psi_{2}|^2 \right) \ee

The result can be changed by inserting a diagonal polarizer into
the path of the recombined beams, in this case the wave function
and probability density will again show interference effects: the
which--path information, connected to the polarization of the two
separate beams is said to have been ''erased''.

\be
\psi_{12}^2 = \frac{1}{2 \sqrt{2}} \left( \psi_{1} + \psi_{2}
\right)\left( \psi_{V} + \psi_{H}\right) \nonumber \\
|\psi_{12}^2|^2 = \frac{1}{4} \left( |\psi_{1}|^2 + |\psi_{2}|^2 +
2 \Re \left[ \psi_{1}^{*} \psi_{2}\right]\right) \ee

In the context of intrinsic properties and polarizations of
intrinsic fields, since the intensity of electromagnetic radiation
is described by the electromagnetic potential $\phi_{em}$ (for a
general description the field vectors are assumed complex):

\be
\phi_{em} = \frac{1}{2} \left(\ \frac{1}{c^2} |\vec E|^2 + |\vec
B|^2 \right) \ee

\noindent If the beam of horizontal polarization (direction $x$)
is split into two separate beams, the electric and magnetic fields
after recombination will be:

\be
\vec E_{12}^0 = \frac{1}{\sqrt{2}} \left( E_{1} \vec e_{x} + E_{2}
\vec e_{x} \right) \nonumber \\ \vec B_{12}^0 = \frac{1}{\sqrt{2}}
\left( B_{1} \vec e_{y} + B_{2} \vec e_{y} \right) \ee

\noindent where the fields $E_{2}, B_{2}$ contain the phase
information $e^{i \varphi}$. The intensity measured after
recombination will consequently contain interference terms:

\be
\phi_{em}^0 = \frac{1}{2} \left( \frac{1}{c^2} |E_{1}|^2 +
|B_{1}|^2 \right) \left(1 + cos \varphi \right) \ee

\noindent A polarization rotator in path 1 changes the
polarization of the electric and magnetic fields to $\vec e_{y}$
and $- \vec e_{x}$ respectively, and the intensity after
recombination is then not affected by the phase $\varphi$:

\be
\vec E_{12}^1 = \frac{1}{\sqrt{2}} \left( E_{1} \vec e_{y} + E_{2}
\vec e_{x} \right) \nonumber \\ \vec B_{12}^1 = \frac{1}{\sqrt{2}}
\left( - B_{1} \vec e_{x} + B_{2} \vec e_{y} \right) \ee

\be
\phi_{em}^1 = \left( \frac{1}{c^2} |E_{1}|^2 + |B_{1}|^2 \right)
\ee

\noindent If the recombined beam is passing through a diagonal
polarizer (plane of polarization in $xy$--direction), the
electromagnetic fields after polarization are:

\be
\vec E_{12}^2 = \frac{1}{2} \left( E_{1} \vec e_{xy} + E_{2} \vec
e_{xy} \right) \nonumber \\ \vec B_{12}^2 = \frac{1}{2} \left( -
B_{1} \vec e_{yx} + B_{2} \vec e_{yx} \right) \ee

\noindent And the intensity of the beam shows again the
interference pattern of the phase $\varphi$:

\be
\phi_{em}^2 = \frac{1}{4} \left( \frac{1}{c^2} |E_{1}|^2 +
|B_{1}|^2 \right) \left(1 + cos \varphi \right) \ee

Mathematically, the description by way of intrinsic potentials and
polarizations yields the same result as the conventional
calculation in quantum theory. However, the interesting aspect of
the effect is its interpretation. While in the conventional
framework the which--path information (and its relation to the
conception of {\em complementarity}) is seen as the ultimate
reason for the experimental results, it is, in the new theory, the
intrinsic information due to the electromagnetic fields and their
vector features, which are held responsible.

\section{Interaction--free measurements}\label{meas_intf}

\begin{figure}
\epsfxsize=1.0\hsize \epsfbox{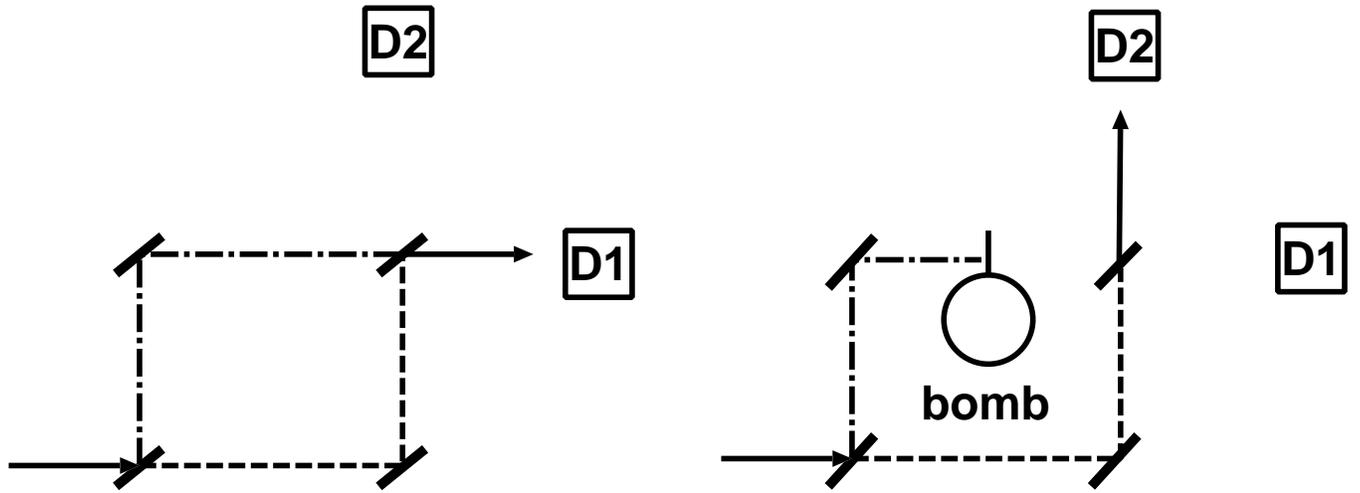} \vspace{0.5cm}
\caption{Mach--Zehnder interferometer with or without a sensitive
bomb--trigger in one path}\label{fig004}
\end{figure}

An interaction--free measurement, which is based on a thought
experiment by Renninger \cite{REN60}, provides information about
the existence of an object in a closed system without necessarily
interacting with this object. The essentials of such a
measurement, recently undertaken by Kwiat et al.
\cite{ZEI95,KWI95}, can be seen in Fig. \ref{fig004}. Interaction
free measurements, usually performed with down--converted photons,
are interesting due to two features: the wave function of the
system and consequently system energy is changed, even if no
interaction occurs. And the results are seemingly incompatible
with classical field theories, because the trajectories of single
particles through the measurement apparatus can be identified.

The first feature was modeled by Dicke \cite{DIC81} using a
modified Heisenberg microscope and calculating the state vectors
of photons and a non--interacting atom in first order perturbation
theory. The result of Dicke's calculation seemed to prove that
even interaction free measurements correlate with an exchange of
virtual photons or, in Dicke's words: {\em The apparent lack of
interaction between the atom and the electromagnetic field is only
illusionary.} On the statistical basis developed in this paper,
the result must be modified. The local modification of ensemble
ranges means, in this context, that an interaction free
measurement corresponds to a different ensemble, i.e. an ensemble
which has zero probability in the range, where an interacting
particle is appreciable. It is therefore the limitation imposed,
the change of boundary conditions, which is the ultimate reason
for the change of the wave function. And if this local range
affects system energy like in Dicke's model of an harmonic
oscillator in a magnetic field \cite{DIC81}, then the energy of
the system changes.

The second feature of interaction free measurements, the assertion
by Kwiat et al. \cite{KWI95} that ''complementary is essential''
to the experimental results achieved, requires a critical
analysis. What the argument indicates, is the impossibility for a
single photon in the interferometer to trigger the bomb {\em and}
detector D2 (see Fig. \ref{fig004}). But as detection efficiency
is only two percent, about 98 \% of the incident energy
(triggering a detector by one of the down--converted photons) is
not accounted for. And in this case the argument of
complementarity as well as the whole argumentation of
interaction--free measurements seems questionable.

\section{Conclusion}

We have shown in this paper that the intrinsic energy component
plays a substantial role in the resolution of some of the most
important paradoxes in quantum theory. In particular we found
that: (i) the spreading of a wave packet is due to a peculiar
choice of initial conditions which, in the present model, are
physically problematic. (ii) The collapse of the wavefunction in
$k$-space is due to a removal of partial waves from the full
ensemble during the measuring process. (iii) The quantum eraser
can be seen as an example, where the polarizations of intrinsic
fields become decisive. (iv) And interaction-free measurements are
not paradoxical, if the local extension of an ensemble is
considered.

\section*{Acknowledgements}

Thanks are due to the \"Osterreichische Forschungsgemeinschaft for
their generous support to attend the Wigner Symposium.

\newpage

\end{document}